\documentclass[conference]{IEEEtran}

\usepackage{amsfonts,amsmath,amssymb,mathrsfs,amsbsy,amsthm}
\usepackage{tikz}
\usepackage{subfigure}
\usepackage{color}
\usepackage{graphicx,epsfig}
\usepackage{stfloats} 
\usepackage[draft]{hyperref}
\usepackage{psfrag}
\usepackage{mathtools}

\newcounter{example}
\newenvironment{example}[1][]{\refstepcounter{example}\par\medskip\noindent%
   \textbf{Example~\theexample.#1} \rmfamily}{\medskip}

\newcounter{const}
\newenvironment{const}[1][]{\refstepcounter{const}\par\medskip\noindent%
   \textbf{Construction~\theconst.#1} \rmfamily}{\medskip}

\newcounter{algorithm}

\newtheoremstyle{mystyle}
  {}
  {}
  {\itshape}
  {}
  {\bfseries}
  {.}
  { }
  {}

\theoremstyle{mystyle}

\newtheorem{thm}{Theorem}
\newtheorem{lem}{Lemma}
\newtheorem{defn}{Definition}

\begin{document}

\sloppy

\title{Binary Cyclic Codes that are Locally Repairable}


\author{
  \IEEEauthorblockN{Sreechakra Goparaju}
  \IEEEauthorblockA{Department of Electrical Engineering\\
    Princeton University\\
    Princeton, NJ, USA\\
    Email: goparaju@princeton.edu}
    \and
  \IEEEauthorblockN{Robert Calderbank}
  \IEEEauthorblockA{Department of Electrical and Computer Engineering\\
    Duke University\\
    Durham, NC, USA\\
    Email: robert.calderbank@duke.edu}
}



\maketitle

\begin{abstract}
Codes for storage systems aim to minimize the repair locality, which is the number of disks (or nodes) that participate in the repair of a single failed disk. Simultaneously, the code must sustain a high rate, operate on a small finite field to be practically significant {\em and} be tolerant to a large number of erasures. To this end, we construct new families of binary linear codes that have an optimal dimension (rate) for a given minimum distance and locality. Specifically, we construct cyclic codes that are locally repairable for locality $2$ and distances $2$, $6$ and $10$. 
In doing so, we discover new upper bounds on the code dimension, and prove the optimality of enabling local repair by provisioning disjoint groups of disks.
Finally, we extend our construction to build codes that have multiple repair sets for each disk. 
\end{abstract}
\section{Introduction}

{\let\thefootnote\relax\footnotetext{The work of Sreechakra Goparaju and Robert Calderbank was supported in part by Air Force Office of Scientific Research grant FA 9550-13-01-0076 under the Complex Networks Program.}}

Triple replication (or {\em triplication}) has been the default storage policy in cloud file systems (notably, Google File system \cite{GGL03} and Hadoop \cite{Hadoop}), and though storage-inefficient, performs extremely well in the face of single disk failures. However, the storage capacity today continues to scale beyond petabytes at a rate steadily incompatible with the adage `{\em storage is cheap}'. Whereas erasure codes such as Reed-Solomon codes are storage-efficient, they make the time to repair a disk failure much longer. The conflicting goals have led the search for new storage codes towards optimizing different repair metrics. A large literature studies {\em regenerating codes}, which minimize the total number of symbols (or bits) communicated in repairing a failed disk, e.g. \cite{DGWWR10, RSK11, SR10, PDC11, CJMRS13}, and \cite{WTB11}. Another metric of importance, and the one we focus on here, is the number of disks (or nodes) which participate in the repair process, referred to as {\em locality}; see among others, \cite{DB08}, \cite{HCL07}, \cite{OD11}, \cite{GHSY12}, \cite{PD12}, \cite{PKLK12}, \cite{TB13}, and \cite{SRKV13}.

\subsection{Locally Repairable Codes} A locally repairable (or recoverable) code (LRC) is a code of length $n$ over a finite field $\mathbb{F}$ such that a symbol in any coordinate\footnote{Each coordinate of a codeword corresponds to a disk or a node in the distributed storage system over $n$ nodes.} of a codeword can be recovered by accessing the symbols in at most $r$ other coordinates. Most of the literature has been devoted to calculating the maximum possible minimum Hamming distance (or just, distance) $d$ achievable for a given code dimension $k$ (or cardinality $A$) and locality $r$. An upper bound on $d$, analogous to the Singleton bound for classical codes, is given by
\begin{eqnarray}\label{eq:singleton}
d &\le& n-k-\left\lceil\frac{k}{r}\right\rceil + 2,
\end{eqnarray}
which has been proved using multiple approaches in \cite{GHSY12, PD12, PKLK12} and \cite{TB13}. LRCs which achieve this distance (constructed, for example, in \cite{SRKV13} and \cite{TPD13}), are called {\em optimal}, and {\em Singleton-optimal} in this paper. Singleton-optimal LRCs were recently constructed \cite{TB13} over a finite field of any size that exceeds the code length $n$ using polynomial evaluations, thereby generalizing Reed-Solomon codes.

In practice, small finite fields, especially Galois fields of the form $\mathbb{F}_{2^m}$, are preferred for their fast arithmetic \cite{PGM13} and backward compatibility with existing hardware. For instance, the recently developed LRC for Microsoft's Windows Azure storage \cite{UsenixLRC12} and that implemented on a Facebook analytics cluster \cite{XorbaFacebook} are both codes over binary extension fields. However, little has been said on explicit LRCs on a given finite field. An upper bound on the distance of an LRC with a given locality $r$ and field size was given in \cite{CM13}, and the family of simplex codes was shown to achieve the bound, thus proving them to be optimal over the binary field, $\mathbb{F}_2$.

\subsection{Contribution}
In this paper, we focus on linear cyclic codes to construct new binary LRCs (that is, LRCs over $\mathbb{F}_2$). Cyclic codes \cite{MacWilliamsSloane} have an inherent structure which enables highly efficient encoder and decoder implementations; they are widely used in data communications and data storage, and several of the well-known classical codes (such as Golay codes, BCH codes and Reed-Solomon codes) are cyclic.

We begin with constructing a family of locally repairable cyclic codes (or, {\em cyclic LRCs}) in Section \ref{sec:singletonOptimal} which turns out to be Singleton-optimal, and acts as a base for the subsequent code constructions. In Section \ref{sec:beyondSingleton}, we construct codes having distances $d = 6$ and $d = 10$ and locality $r = 2$. For each of these families, we analyze (and in most cases, prove) their optimality in terms of their dimension. 

The structure of cyclic codes seamlessly leads to a division of codeword coordinates into disjoint repair groups, such that the coordinates in each repair group aid in the repair of each other (and do not need other coordinates). This structural assumption is either part of the repair model or proved to be optimal in several cases, e.g. \cite{PKLK12, GHSY12}, and is useful in constructing LRCs over small finite fields. In particular, it has been proved \cite{TB13} to be optimal when $r+1$ divides $n$ and the field size $|\mathbb{F}|\ge n$. We prove that the optimality of this assumption continues to hold for a family of codes constructed in Section \ref{sec:beyondSingleton}. 

The structural assumption also leads to the {\em tensor-product} style of construction that we use and has been alluded to in \cite{TB13} and \cite{CM13}.
In Section \ref{sec:availability}, we extend this tensor-product construction to obtain codes with multiple {\em availability}, wherein every symbol can be simultaneously repaired by more than one repair set of size $r$. Availability is a useful property in systems storing frequently-accessed data; see \cite{PHO13, TB13}, and \cite{RPDV13} for some distance optimality results and multi-available LRC constructions.
{\color{black}Finally, we conclude in Section \ref{sec:conclusion}.}

\section{A Singleton-Optimal Family}\label{sec:singletonOptimal}
We first redefine an LRC in terms of binary linear codes.
\begin{defn}[Locally Repairable Code]\label{defn:LRC}
A binary linear {\em locally repairable code (LRC)} of locality $r$ is defined as a binary linear code $\mathscr{C}$ of length $n$, 
such that every coordinate in $[n]:=\{1,\ldots,n\}$ is contained in the support of some parity check of weight $r+1$.
\end{defn}

For simplicity we focus on construction of
locally repairable cyclic codes over $\mathbb{F}_2$ of length
$n=2^m-1$,
and locality $r$ that is even.
Furthermore, as mentioned earlier, we assume
that $r+1 \,|\, n$. 
It follows that $m$ is even when $r=2$.

\subsection{Motivation}
Consider the linear code dual to the LRC we intend to construct. The smallest cyclic code containing codewords of (Hamming) weight $r+1$ is that which consists of the cyclic shifts of the codeword with ones in positions $0, n/(r+1), 2n/(r+1)$, and $rn/(r+1)$. This code exists. If $\alpha$ is a primitive element\footnote{Throughout the paper, $\alpha$ will represent a primitive element of $\mathbb{F}_{2^m}$.} (representable as a binary $m$-vector), then we have $\alpha$ as a zero of
\begin{eqnarray*}
h(x) &=& \sum_{i=0}^r x^{in/(r+1)},
\end{eqnarray*} 
where the polynomial $h(x)$ generates the dual code. Notice that if $g(x)$ and $h(x)$ are the generator polynomials of the LRC and its dual, respectively, then
\begin{eqnarray*}
x^n-1 &=& g(x)h(x),
\end{eqnarray*}
and every non-zero field element $\alpha^j \in \mathbb{F}_{2^m}$, where $j \in \{0,1,\ldots,n-2\}$, is a root of either $h(x)$ or $g(x)$. It can be verified then that $h(x)$ is a factor of $x^n-1$ and that the other zeroes\footnote{It is here that we use that $r$ is even. We do not want $h(\alpha^0) = 0$.} of $h(x)$ are $\alpha^{j}$, such that $j$ is not divisible by $r+1$. We therefore have the following family of cyclic LRCs.

\begin{const}\label{const:1}
Let $n=2^m-1$, $r+1$ be a factor of $n$ and $\alpha$ be a primitive element of $\mathbb{F}_{2^m}$. Let $\mathscr{C}$ be a cyclic code with the generator polynomial $g(x)$ having zeroes $\alpha^{j(r+1)}$, where $j$ ranges from $0$ to $(n/(r+1))-1$. Then $\mathscr{C}$ is an LRC with locality $r$, and dimension $k := rn/(r+1)$. 
\end{const}

\begin{example}\label{ex:15-10-2-2-1}
For $n = 15$ and locality $r = 2$, we have the following codewords in the dual code
\begin{eqnarray*}
&&{\sf 1\,\,0\,\,0\,\,0\,\,0\,\,1\,\,0\,\,0\,\,0\,\,0\,\,1\,\,0\,\,0\,\,0\,\,0},\\
&&{\sf 0\,\,1\,\,0\,\,0\,\,0\,\,0\,\,1\,\,0\,\,0\,\,0\,\,0\,\,1\,\,0\,\,0\,\,0},\\
&&{\sf 0\,\,0\,\,1\,\,0\,\,0\,\,0\,\,0\,\,1\,\,0\,\,0\,\,0\,\,0\,\,1\,\,0\,\,0},\\
&&{\sf 0\,\,0\,\,0\,\,1\,\,0\,\,0\,\,0\,\,0\,\,1\,\,0\,\,0\,\,0\,\,0\,\,1\,\,0},\\
&&{\sf 0\,\,0\,\,0\,\,0\,\,1\,\,0\,\,0\,\,0\,\,0\,\,1\,\,0\,\,0\,\,0\,\,0\,\,1},
\end{eqnarray*}
which leads to the following {\em check} polynomial,
\begin{eqnarray*}
h(x) &=& 1+x^5+x^{10},\\
&=& (x-\alpha)(x-\alpha^2)(x-\alpha^4)(x-\alpha^5)\cdots(x-\alpha^{14}),
\end{eqnarray*}
and an LRC of dimension $k = 10$.
\end{example}

\smallskip
Using the LRC Singleton bound from (\ref{eq:singleton}),
we have,
\begin{eqnarray*}
d &\le& n - \frac{rn}{r+1} - \left\lceil\frac{rn}{r(r+1)}\right\rceil+2\\
&=& 2.
\end{eqnarray*}
This upper bound on $d$ is achievable for the code defined in Construction \ref{const:1} (a codeword of Hamming weight $2$ can be found -- for example, ones at positions $0$ and $n/(r+1)$), and therefore Construction \ref{const:1} gives a family of Singleton-optimal codes\footnote{These codes are intolerant to multiple failures but efficiently repair single failures.}. As in classical cyclic codes, this minimum distance can be increased by adding more zeroes to the generator polynomial. We look at the optimality of some such codes in the next section.

\section{Beyond Singleton-Optimality}\label{sec:beyondSingleton}

\subsection{Disjoint Repair Groups}

We first construct and analyze some codes which have disjoint repair groups. Consider the following code with minimum distance $d = 6$.
\begin{example}
For $n = 15$ and locality $r = 2$, if we have the zeroes of $g(x)$ as $1$ and all cyclotomic cosets of $\alpha$ and $\alpha^3$, then $k = 6$ and $d = 6$.
\end{example}

This is not optimal with respect to the LRC Singleton bound (\ref{eq:singleton}), which gives $d \le 8$. In general, we have the following family of codes.

\begin{const}\label{const:2}
Let $n = 2^m-1$ with $m$ even, and locality $r=2$. Let $\mathscr{C}$ be a cyclic code with the generator polynomial $g(x)$ having zeroes as in Construction \ref{const:1}, along with the cyclotomic coset of $\alpha$. Then $\mathscr{C}$ is an LRC of dimension
\begin{eqnarray*}
k &=& \frac{2}{3}\left(2^m-1\right)-m,
\end{eqnarray*}
and a distance $d \ge 6$.
\end{const}

The lower bound on $d$ follows from the BCH distance bound (e.g., \cite{MacWilliamsSloane}). From the LRC Singleton bound (\ref{eq:singleton}), we have
\begin{eqnarray*}
d &\le& \frac{3m}{2} + 2.
\end{eqnarray*}

However, we show in the next theorem that this family of codes is indeed distance-optimal among the set of linear codes which have disjoint locality parity checks.

\begin{thm}\label{thm:d6r2}
Consider a binary linear code $\mathscr{C}$ of length $n = 2^m-1$, distance $d = 6$, and locality $r = 2$. Let $m$ be an even number greater than $2$. Suppose that the set of coordinates $[n]$ can be divided into $n/3$ groups, $\{g_i\}_{i=1}^{n/3}$, such that the repair of a given coordinate only requires the bits stored in the coordinates in its corresponding group. This implies that the dual code has a codeword (parity-check) of Hamming weight $r+1=3$ supported by the coordinates in each group $g_i$. Then,
\begin{eqnarray}\label{eq:d6r2}
k &\le& \frac{2}{3}\left(2^m-1\right)-m.
\end{eqnarray}
Notice that this $k$ corresponds to the dimension of the family of codes in Construction \ref{const:2}.
\end{thm}
\begin{IEEEproof}
Let ${\bf x} = \left(x_1, x_2, \ldots, x_n\right)$ be a codeword in $\mathscr{C}$, and let $g_i = \{i,i+(n/3),i+(2n/3)\}$ for $i = [n/3]$. Consider the projection of a codeword ${\bf x}$ onto the set of coordinates $g_i$. By the locality parity-checks, we know that the only possible projections are $\{{\sf 000,011,101,110}\}$. In other words, we can map the binary code $\mathscr{C}$ to an {\em additive} code $\mathscr{C}_4$ over $\mathbb{F}_4$ of length $n/3$, minimum distance $d/2$ and size $2^k$. Conversely, any additive code $\mathscr{C}_4$ over $\mathbb{F}_4$ of length $n'=(2^m-1)/3$ and minimum (Hamming) distance $d'=3$ maps back to a binary code $\mathscr{C}$ which satisfies the theorem's conditions. Thus, we need only prove that the dimension for such an additive code is upper bounded by
\begin{eqnarray}\label{eq:ubGF4}
k' &\le& \frac{2^m-1}{3}-\frac{m}{2}.
\end{eqnarray}

To this end, consider the given code $\mathscr{C}_4$ and let ${\bf x'} = (x'_1, \ldots, x'_{n'})$ be a codeword in $\mathscr{C}_4$. We will prove that there exists a set $M$ of $m/2$ coordinates in $[n']$ such that the support of no non-zero codeword is a subset of $M$, that is, $\mathsf{supp}({\bf x'}) \not \subseteq M$, for all ${\bf x'} \in \mathscr{C}_4\backslash \{{\bf 0}\}$. Then, the code $\mathscr{C}_4$ projected onto the rest of the coordinates, has the same dimension as $\mathscr{C}_4$, that is, the projected code of length $n'-(m/2)$ has a minimum (Hamming) distance of at least $1$. Thus, from the na\"{i}ve Singleton bound applied to the projected code, we have
\begin{eqnarray}
k' &\le& n'-\frac{m}{2},
\end{eqnarray}
which proves (\ref{eq:ubGF4}).

 Let $u$ be any nonnegative integer\footnote{We can assume that $m \ge 6$ here; for $m=4$, any $M$ is sufficient.} such that 
 \begin{eqnarray}\label{eq:theCondition}
 4^u &<& n'-u.
\end{eqnarray}
Now consider the collection of $n'-u$ sets of coordinates of the form $\{i_1,i_2,\ldots,i_u,i_{u+1}\}$, where the first $u$ coordinates are fixed and $i_{u+1} \in [n']\backslash\{i_1,\ldots,i_u\}$. We claim that there exists at least one set in this collection such that there is no codeword ${\bf x'}$ such that $i_{u+1} \in \mathsf{supp}({\bf x'}) \subseteq \{i_1,\ldots,i_{u+1}\}$. If this were not true, then there exists a unique codeword corresponding to each of the $n'-u$. From (\ref{eq:theCondition}), two of these codewords have the same projections on the coordinates $\{i_1, \ldots, i_u\}$ and therefore, are at a distance of $2$, a contradiction. We can now use this claim to construct the required $M$.

Without loss of generality, let $i_1 = 1$ and $i_2 = 2$. Since $4^2 < n'-2$ for any $m > 4$, there exists a set of coordinates $\{1,2,i_3\}$ such that there is no codeword ${\bf x'}$ which satisfies $i_{3} \in \mathsf{supp}({\bf x'}) \subseteq \{1,2,i_{3}\}$ (in this special case, we just mean that there is no codeword with the support $\{1,2,i_3\}$). Again, with no loss in generality, let $i_3 = 3$. We can append $i_4=4$ to the set $\{1,2,3\}$ in the same manner as above if $u=4$ satisfies (\ref{eq:theCondition}). There is then no codeword ${\bf x'}$ such that $4 \in \mathsf{supp}({\bf x'}) \subseteq \{1,2,3,4\}$. Continuing the process, we obtain a set of coordinates  $\{1,\ldots,u^{*},u^{*}+1\}$ such that there is no codeword ${\bf x'}$ which satisfies any of the following conditions:
\begin{eqnarray*}
\begin{array}{ccccc}
u+1 &\in& \mathsf{supp}({\bf x'}) &\subseteq& \{1,2,\ldots,u,u+1\},
\end{array}
\end{eqnarray*}
where $u \in \{2,3,\ldots,u^{*}\}$ and $u^{*}$ is the maximum value of $u$ which satisfies (\ref{eq:theCondition}). It can be verified that the above construction implies that for all ${\bf x'} \in \mathscr{C}\backslash \{{\bf 0}\}$, $\mathsf{supp}({\bf x'}) \not \subseteq [u^{*}+1]$, and that $u^{*} = m/2-1$ for the given parameters. Thus, $M = [u^{*}+1]$ has been constructed.
\end{IEEEproof}
\begin{IEEEproof}[{\color{black}Another Proof}]
The upper bound in (\ref{eq:ubGF4}) follows directly from the Hamming bound. In fact, it is satisfied with equality by the Hamming codes over $\mathbb{F}_4$ with the parameters
\begin{eqnarray*}
\begin{array}{ccccccccc}
n' &=& \displaystyle\frac{4^{m/2}-1}{4-1},&k'&=&n'-\displaystyle\frac{m}{2},&d'&=&3.
\end{array}
\end{eqnarray*}

In other words, the constructed binary $2$-local codes map to the Hamming codes over $\mathbb{F}_4$.
\end{IEEEproof}

\begin{thm}\label{thm:d10r2}
Consider a binary linear code $\mathscr{C}$ of length $n = 2^m-1$, minimum distance $d = 10$, and locality $r = 2$. Let $m$ be an even number greater than $2$. Suppose that the set of coordinates $[n]:=\{1,\ldots,n\}$ can be divided into disjoint repair groups as in Theorem \ref{thm:d6r2}. Then,
\begin{eqnarray*}
k &\le& \frac{2}{3}\left(2^m-1\right)-2m + 1.
\end{eqnarray*}
However, if $k$ is even, then
\begin{eqnarray}\label{eq:linearGF4Bound}
k &\le& \frac{2}{3}\left(2^m-1\right)-2m,
\end{eqnarray}
and there exists a family of cyclic codes satisfying this bound with equality.
\end{thm}
\begin{IEEEproof}
The proof follows in a similar way as that of Theorem \ref{thm:d6r2}. The binary code $\mathscr{C}$ can be mapped to an additive code $\mathscr{C}_4$ over $\mathbb{F}_4$ of length $n'=n/3$, minimum distance $d'=5$ and size $2^k$. By the Hamming bound, we have the maximum size $A(n',5) := A$ of any code over $\mathbb{F}_4$ of length $n'$ and $d'=5$ upper bounded as
\begin{eqnarray}\label{eq:ubHamming}
\begin{array}{ccccc}
2^k &\le& A &\le& \displaystyle{\frac{4^{n'}}{1+3n'+\displaystyle\frac{9n'(n'-1)}{2}}},
\end{array}
\end{eqnarray}
and the substitution $n = 2^m-1$ leads to
\begin{eqnarray}
k &\le& \left\lfloor\frac{2n}{3}+1-\log_2\left(2^{2m}-3(2^m)+4\right)\right\rfloor,\nonumber\\
&=& \frac{2n}{3}+1-\left\lceil\log_2\left(2^{2m}-3(2^m)+4\right)\right\rceil,\nonumber\\
&=& \displaystyle{\frac{2n}{3}+1-2m},\label{eq:additiveBound}
\end{eqnarray}
for any even $m > 2$. Moreover, if $k$ is even, we have (\ref{eq:linearGF4Bound}) instead.

{\em Remark:} If we assume that the contracted code is linear, then $k$ is always even (so that the dimension of the code $\mathscr{C}_4$ over $\mathbb{F}_4$ is an integer). This gives us the required bound in (\ref{eq:linearGF4Bound}).

\begin{const}The upper bound (\ref{eq:linearGF4Bound}) is achievable by a cyclic code whose generator polynomial $g(x)$ has zeros at $\alpha^j$, where $j$ is a multiple of $3$; $\alpha^{2^i}$, that is, the cyclotomic cosets of $\alpha$; and $\alpha^{n-2^i}$, that is, the cyclotomic cosets of $\alpha^{-1}$, where $\alpha$ is a primitive element.
\end{const}
\end{IEEEproof}

\subsection{Optimality of Disjoint Repair Groups}
We now prove that the assumption of disjoint locality parity checks leads to no loss in optimality in Theorem \ref{thm:d6r2}. We start with a basic lemma on the locality parity checks.
\begin{lem}\label{lem:linIndOfPCs}
Let $\mathscr{C}$ be a binary linear code of length $n$ and locality $r$. Then there exist linearly independent codewords (parity checks) of weight $r+1$ in the dual code $\mathscr{C}^{\perp}$, the union of whose supports equals $[n]$.
\end{lem}
\begin{IEEEproof}
Let ${\cal P}$ be the set of parity checks corresponding to every coordinate in a codeword of $\mathscr{C}$, and let ${\cal P}_{{\sf m}}$ be a maximal linearly independent subset of ${\cal P}$. If the supports of the parity checks in ${\cal P}_{{\sf m}}$ do not cover a coordinate in $[n]$, say $i$, then any parity check in ${\cal P}$ corresponding to $i$ lies outside the span of ${\cal P}_{{\sf m}}$, disproving its maximality. Thus, ${\cal P}_{{\sf m}}$ satisfies the lemma.

{\em Remark:} In the next theorem, we use not ${\cal P}_{{\sf m}}$, but the smallest subset of ${\cal P}_{{\sf m}}$ which satisfies the lemma.
\end{IEEEproof}


\begin{thm}\label{thm:d6r2StructureTheorem}
Consider a binary linear code $\mathscr{C}$ of length $n = 2^m-1$, distance $d=6$, and locality $r=2$. If $2\,|\,m$ and $m > 8$, then the upper bound {\em(\ref{eq:d6r2})} on the dimension $k$ of $\mathscr{C}$, 
\begin{eqnarray*}
k &\le& \frac{2}{3}\left(2^m-1\right)-m,
\end{eqnarray*}
continues to hold.
\end{thm}
\begin{IEEEproof}
Let ${\cal Q}$ be a set of 
linearly independent parity checks of weight $3$ in $\mathscr{C}^{\perp}$ that cover $[n]$. Suppose that its cardinality, $|{\cal Q}|=b$, be minimal and 
be given by
\begin{eqnarray}\label{eq:linIndParities}
b &=& \frac{2^m-1}{3} + t,
\end{eqnarray}
where $0\le t \le m$. Here $t \ge 0$ because we need at least $n/3$ repair parity checks to cover each coordinate. For $t \ge m+1$, we have
\vspace{0.1cm}
\begin{eqnarray*}
\begin{array}{ccccl}
n-k &\ge& b &\ge& \displaystyle\frac{2^m-1}{3}+(m+1), \,\,\,
\vspace{0.3cm}\textrm{that is},\\
&&k &\le& \displaystyle\frac{2}{3}\left(2^m-1\right)-m-1,
\end{array}
\end{eqnarray*}
which satisfies (\ref{eq:d6r2}), with a strict inequality.

Let $N$ be the maximum number of pairwise disjoint weight $3$ parity checks in ${\cal P}_{{\sf m}}$. These parity checks cover $3N$ coordinates, and each of the remaining $b-N$ covers at most $2$ additional coordinates (not already covered by the $N$ in the disjoint set), that is,
\begin{eqnarray*}
3N+2(b-N) &\ge& n,\,\,\,\textrm{that is},\\[0.1cm]
N &\ge& \frac{2^m-1}{3} - 2t,\\[0.2cm]
&\ge& \frac{2^m-1}{3}-2m.
\end{eqnarray*}
The remaining $b-N \le 3t \le 3m$ parity checks can overlap with at most $6m$ of the $N$ pairwise disjoint parity checks (if each overlaps with two different parity checks). We therefore have at least $N^{*}$ pairwise disjoint parity checks that do not intersect with any of the remaining $b-N^{*}$ parity checks in ${\cal Q}$, where
\begin{eqnarray*}
N^{*} &\ge& \frac{2^m-1}{3} - 8m.
\end{eqnarray*} 
Notice that the right hand side above is positive for $m \ge 8$. Let ${\cal N}$ be the set of $3N^{*}$ coordinates covered by these $N^{*}$ parity checks. 
Consider the sub-code $\mathscr{C}_{{\cal N}}$ of $\mathscr{C}$, with zeroes in the coordinates $[n]\backslash {\cal N}$. The $b$ parity checks in ${\cal Q}$ do not preclude codewords which have weights $2$ and $4$ in $\mathscr{C}_{\cal N}$. To impose a distance $d = 6$ on $\mathscr{C}_{{\cal N}}$, the codewords in $\mathscr{C}_{{\cal N}}$ of weight less than $3$ (that is, of weights $0$ and $2$) must be in different cosets, that is,
\begin{eqnarray*}
2^{\hat{b}} &\ge& 1 + 3N^{*},\\
&\ge& 2^m-24m,
\end{eqnarray*}
where $\hat{b}$ is the number of additional parity checks necessary. Note that the right hand side can be viewed as the number of codewords of length $N^{*}$ on $\mathbb{F}_4$ of weight less than $2$ (with a mapping similar to that in the proof of Theorem \ref{thm:d6r2}). We therefore have
$\hat{b} \ge m$,
for $m > 8$. This, coupled with (\ref{eq:linIndParities}), gives the dimension of $\mathscr{C}^{\perp}$ as
\begin{eqnarray*}
n-k &\ge& b + \hat{b},\\[0.1cm]
&\ge& \frac{2^m-1}{3} + m,
\end{eqnarray*}
which is the same as (\ref{eq:d6r2}).
\end{IEEEproof}
\smallskip
{\em Remark:} 
If $t=0$, then Theorem \ref{thm:d6r2} applies, and if $t>0$, then
 we have $n-k \ge (n/3)+m+1$, which satisfies (\ref{eq:d6r2}) with a strict inequality.

\begin{thm}[Corollary]
Let $\mathscr{C}$ be a binary linear code as given in {\em Theorem \ref{thm:d6r2StructureTheorem}}. Then, $\mathscr{C}$ is a distance-optimal LRC only if it has disjoint locality parity checks.
\end{thm}


\section{Multiple Repair Sets}\label{sec:availability}
The basic cyclic code construction in sections \ref{sec:singletonOptimal} and \ref{sec:beyondSingleton} (Construction \ref{const:1}) can be seen as a linear code with a parity check matrix given by the tensor product 
\begin{eqnarray*}
\left({\sf 1\,\,1\,\,1\,\,\cdots\,\,1}\right) &\otimes&  {\bf I}_{n/(r+1)} ,
\end{eqnarray*}
where the first matrix, having $(r+1)$ ones is the parity check matrix for the simplest binary LRC with locality $r$ and ${\bf I}_{n/(r+1)}$ is an $n/(r+1)$-dimensional identity matrix. This construction can similarly be extended to the case of what we call {\em multiply available locally repairable codes} or {\em $t$-available-$r$-local} LRCs. The codes considered in the previous sections were $1$-available-$r$-local LRCs.
\begin{defn}[Availability]\label{defn:availability}
A binary linear code $\mathscr{C}$ of length $n$ is called a {\em $t$-available-$r$-local locally repairable code (LRC)} if every coordinate $i$ in $[n]$ has at least $t$ parity checks of weight $r+1$ which intersect pairwise in (and only in) $\{i\}$.
\end{defn}

In the interest of space, we present only an example of a $3$-available-$2$-local LRC. 
\begin{example}
Consider a binary linear code $\mathscr{C}$ of length $63$ with a parity check matrix given by
\begin{eqnarray*}
H_{{\sf [7,4,3]}} &\otimes&  {\bf I}_{9} ,
\end{eqnarray*}
where $H_{{\sf [7,4,3]}}$ is the parity check matrix of a Hamming $[7,4,3]$ code. It can be verified that the $[7,4,3]$ code is $3$-available-$2$-local, and so $\mathscr{C}$, which is a tensor product of the dual $[7,3,4]$ code and ${\bf I}_{9}$, is a $[n=63,k=27,d=4]$ $3$-available-$2$-local LRC. A generalizing cyclic code construction follows.
\end{example}
\begin{const}
Let $n = 2^m-1$ be divisible by $7$ (that is, $3\,|\,m$). Let $\mathscr{C}$ be a cyclic code with the generator polynomial $g(x)$ having zeroes $\alpha^{j}$, where $j \in \{0,1\ldots,n-1\}$ and $j \,({\sf mod}\,\, 7) \in \{0,3,5,6\}$. Then $\mathscr{C}$ is a $3$-available-$2$-local LRC with dimension $k = 3n/7$, and distance $d = 4$. The check polynomial $h(x)$ is given by
\begin{eqnarray*}
h(x) &=& 1+x^{n/7}+x^{3n/7},
\end{eqnarray*}
which is the generator polynomial in $y := x^{n/7}$ for the Hamming $[7,4,3]$ code. Notice that this construction can be viewed as an additive code over $\mathbb{F}_8$, in an analogous manner to the LRC in Theorem \ref{thm:d6r2}.
\end{const}

\section{Concluding Remarks}\label{sec:conclusion}
We have presented a method of constructing locally repairable codes that preserves the simplicity of three way replication and essentially doubles the data rate. We have also proved that in some cases the data rate of the new codes is optimal. Our construction is a natural extension of historical methods for constructing algebraic error correcting codes used in disk arrays. It leads to new coding theory questions about optimality of binary codes given a constraint on the weight structure of the dual code.

\section{Acknowledgement}
We thank Dimitris Papailiopoulos and Itzhak Tamo for introducing us to this problem, and Itzhak Tamo and Alexander Barg for sharing their preprint \cite{TB13}.

%
%

\bibliographystyle{IEEEtran}
\bibliography{IEEEabrv_sg,references}

\end{document}